\documentclass{aastex}
\usepackage{emulateapj5}

\shorttitle{CHANDRA X-RAY OBSERVATIONS OF NGC~4697}
\shortauthors{SARAZIN, IRWIN, \& BREGMAN}

\slugcomment{Astrophysical Journal Letters, in press}

\begin{document}

\title{Resolving the Mystery of X-ray Faint Elliptical Galaxies:
Chandra X-ray Observations of NGC~4697}

\author{Craig L. Sarazin\altaffilmark{1},
Jimmy A. Irwin\altaffilmark{2,3},
and Joel N. Bregman\altaffilmark{2}}

\altaffiltext{1}{Department of Astronomy, University of Virginia,
P. O. Box 3818, Charlottesville, VA 22903-0818;
cls7i@virginia.edu}

\altaffiltext{2}{Department of Astronomy, University of Michigan,
Ann Arbor, MI 48109-1090; jirwin@astro.lsa.umich.edu, jbregman@umich.edu}

\altaffiltext{3}{Chandra Fellow}

\begin{abstract}
{\it Chandra} observations of the X-ray faint
elliptical galaxy NGC~4697 resolve much of the X-ray emission
(61\% within one effective radius) into $\sim$80 point sources,
of which most are low mass X-ray binaries (LMXBs).
These LMXBs provide the bulk of the hard emission and much of the soft
emission as well.
Of the remaining unresolved emission, it is likely that about half is
from fainter LMXBs, while the other half ($\sim$23\% of the total emission)
is from interstellar gas.
Three of the resolved sources are supersoft sources.
In the outer regions of NGC~4697, eight of the LMXBs (about 25\%)
are coincident with candidate globular clusters,
which indicates that globulars have a high probability of containing
X-ray binaries compared to the normal stellar population.
The X-ray luminosities (0.3--10 keV) of the resolved LMXBs range from
$\sim$$5 \times 10^{37}$ to $\sim$$2.5 \times 10^{39}$ ergs s$^{-1}$.
The luminosity function of the LMXBs has a ``knee'' at
$3.2 \times 10^{38}$ ergs s$^{-1}$, which is roughly the Eddington
luminosity of a 1.4 $M_\odot$ neutron star (NS);
this knee might be useful as a distance indicator.
The highest luminosity source has the Eddington luminosity of
a $\sim$20 $M_\odot$ black hole (BH).
The presence of this large population of NS and massive BH stellar remnants
in this elliptical galaxy shows that it (or its progenitors) once contained
a large population of massive main sequence stars.
\end{abstract}


\section{Introduction} \label{sec:intro}

Observations with the {\it Einstein} X-ray Observatory showed that
elliptical and S0 galaxies are luminous sources of X-ray emission
(e.g., Forman, Jones, \& Tucker 1985).
At least for the X-ray luminous early-type galaxies (defined as
those having a relatively high ratio of X-ray to optical luminosity
$L_X/L_B$),
the bulk of the X-ray luminosity is from hot
($\sim$$10^7$ K) interstellar gas.
However, there is a large dispersion in the X-ray luminosities of
early-type galaxies of a given optical luminosity.
We will refer to galaxies which have a very low $L_X/L_B$ ratio
as ``X-ray faint.''
In these X-ray faint ellipticals, much of the hot interstellar gas
may have been lost in galactic winds
or by ram pressure stripping by ambient intracluster or intragroup gas.
The source of the bulk of the X-ray emission in X-ray faint galaxies
is uncertain.

In general, X-ray faint galaxies exhibit significantly different
spectral properties than their X-ray bright counterparts.
The X-ray bright galaxies are dominated by thermal emission at
$kT \sim 0.8$ keV due to their hot interstellar medium (ISM).
On the other hand, the X-ray faint galaxies exhibit two distinct spectral
components:
first, they have a hard $\sim$5--10 keV component,
most easily seen in {\it ASCA} spectra
(Matsumoto et al.\ 1997),
which is roughly proportional to the optical luminosity of the galaxy.
This suggests that the hard component is due to low-mass X-ray binaries
(LMXBs) like those seen in the bulge of our Galaxy and M31.
X-ray faint galaxies also have
a very soft ($\sim$0.2 keV) component, whose origin is uncertain
(Fabbiano, Kim, \& Trinchieri 1994; Pellegrini 1994; Kim et al.\ 1996).
Suggested stellar sources for the very soft emission in X-ray faint ellipticals
include active M stars, RS CVn binaries, or supersoft sources, but none of
these appears to work quantitatively
(Pellegrini \& Fabbiano 1994; Irwin \& Sarazin 1998b).
It is possible that the soft X-rays are due to warm (0.2 keV) ISM.
Recently, we proposed that the very soft emission in X-ray
faint ellipticals is due to the same LMXBs responsible for the hard emission
(Irwin \& Sarazin 1998a,b).
However, the origin of the very soft component in X-ray faint ellipticals
remains something of a mystery.

With the superb spatial resolution of the {\it Chandra} X-ray
Observatory, it should be possible to resolve the emission from
nearby X-ray faint early-type galaxies into LMXBs, if this is
indeed the source of their emission.
Recently, we used a deep {\it ROSAT} HRI observation to detect a
number of discrete X-ray sources in the X-ray faint elliptical NGC~4697
(Irwin, Sarazin, \& Bregman 2000, hereafter ISB).
However, the bulk of the X-ray emission was not resolved.
We also simulated a 40 ksec {\it Chandra} observation of NGC~4697,
and showed that it should be possible to detect $\sim$100 LMXBs
if they provide the bulk of the emission.
Here, we present the results of exactly this observation.
At a distance of 15.9 Mpc
(Faber et. al.\ 1989; assuming a Hubble constant
of 50 km s$^{-1}$ Mpc$^{-1}$),
NGC~4697 is the closest normal, optically luminous, X-ray
faint elliptical galaxy.
Given its proximity, NGC~4697 is an ideal target for detecting
the LMXB population.
It is sufficiently X-ray faint that diffuse ISM emission should not 
bury the fainter LMXBs.
The purposes of the {\it Chandra} observation are
to resolve and study the LMXB population of NGC~4697,
to determine the source of both the hard and soft spectral components,
and
to detect or place strong limits on any residual diffuse (possibly
gaseous) emission.

\section{X-ray Observation} \label{sec:obs}

NGC~4697 was observed on 2000 January 15-16
with the {\it Chandra} ACIS-S3 detector for a total exposure of 39,434 s.
The S3 chip was chosen because of its good low energy response.
No background flares occurred during this observation.
This observation was processed at a time when the standard pipeline
processing introduced a boresight error of about 8\arcsec\ in the
absolute positions of X-ray sources.
The absolute positions were corrected by using the optical identifications and
positions from USNO-A2.0 (Monet et al.\ 1998) of X-ray sources.
The absolute positions are accurate
to about 0.5\arcsec\ near the center of the S3 image, with larger errors
further out.
Background was taken from the outer regions of the same chip and
blank sky backgrounds.
More details concerning the X-ray observation, data analysis, variability
of the sources, and spectral fitting are given in
Sarazin, Irwin, \& Bregman (2000, hereafter Paper II).

%
%
\centerline{\null}
\vskip2.75truein
\includegraphics{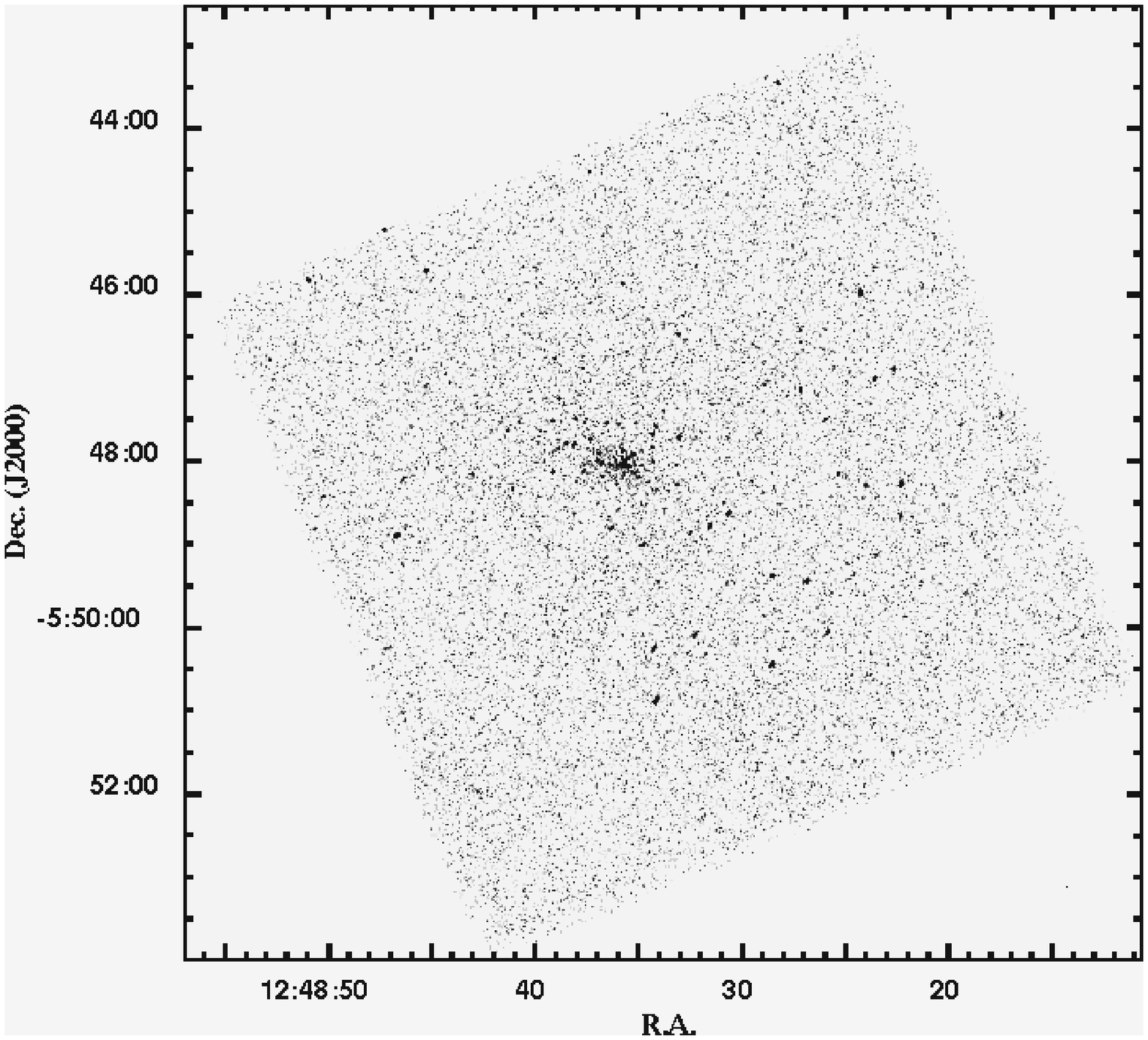}
\figcaption{
The entire {\it Chandra} S3 X-ray image of the region around
NGC~4697 in the 0.3--10 keV band.
The optical center of the galaxy is located at the center of the
concentration of X-ray point sources to the northeast of the chip center.
\label{fig:xray_whole}}

\vskip0.2truein

\section{X-ray Image} \label{sec:image}

The {\it Chandra} S3 chip X-ray image is shown in
Figure~\ref{fig:xray_whole}.
The central portion of the {\it Chandra} image (adaptively smoothed)
is shown in Figure~\ref{fig:xray_mid}.
These images show the basic result of the {\it Chandra} observation:
much of the emission from the galaxy is resolved into individual point
sources of X-rays (\S~\ref{sec:src}).
However, Figure~\ref{fig:xray_mid} indicates that there is also some
unresolved, diffuse emission.
The diffuse emission is centrally concentrated, and its
X-ray surface brightness increases by about a factor of 30 
into the center of the galaxy.
Both the point sources and the diffuse emission are elongated in the
same direction as this E6 galaxy
(a position angle of 67$^\circ$).

We determined the portion of the X-ray emission due to resolved
sources and diffuse emission, both for the entire {\it Chandra }
X-ray band (0.3--10 keV) and for three narrower bands:
hard H (2-10 keV), medium M (1-2 keV), and soft S (0.3-1 keV).
Here, we report on the counts coming from within the elliptical optical
isophote containing one half of the optical light
(``one effective radius,'' semimajor axis $95^{\prime\prime}$,
semiminor axis $55^{\prime\prime}$).
First, the population of resolved sources was determined as discussed
below (\S~\ref{sec:src}).
Then, the total emission was determined, and the source flux subtracted
to give the amount of unresolved emission.

Within one effective radius, 61\% of the X-ray emission is resolved
into individual X-ray sources for the total band.
In the hard, medium, and soft bands, the resolved fractions are
81\%, 79\%, and 48\%, respectively.
Thus, it is clear that the bulk of the total X-ray emission from NGC~4697
comes from point sources, which are likely to be LMXBs.

%
%
\centerline{\null}
\vskip2.75truein
\includegraphics{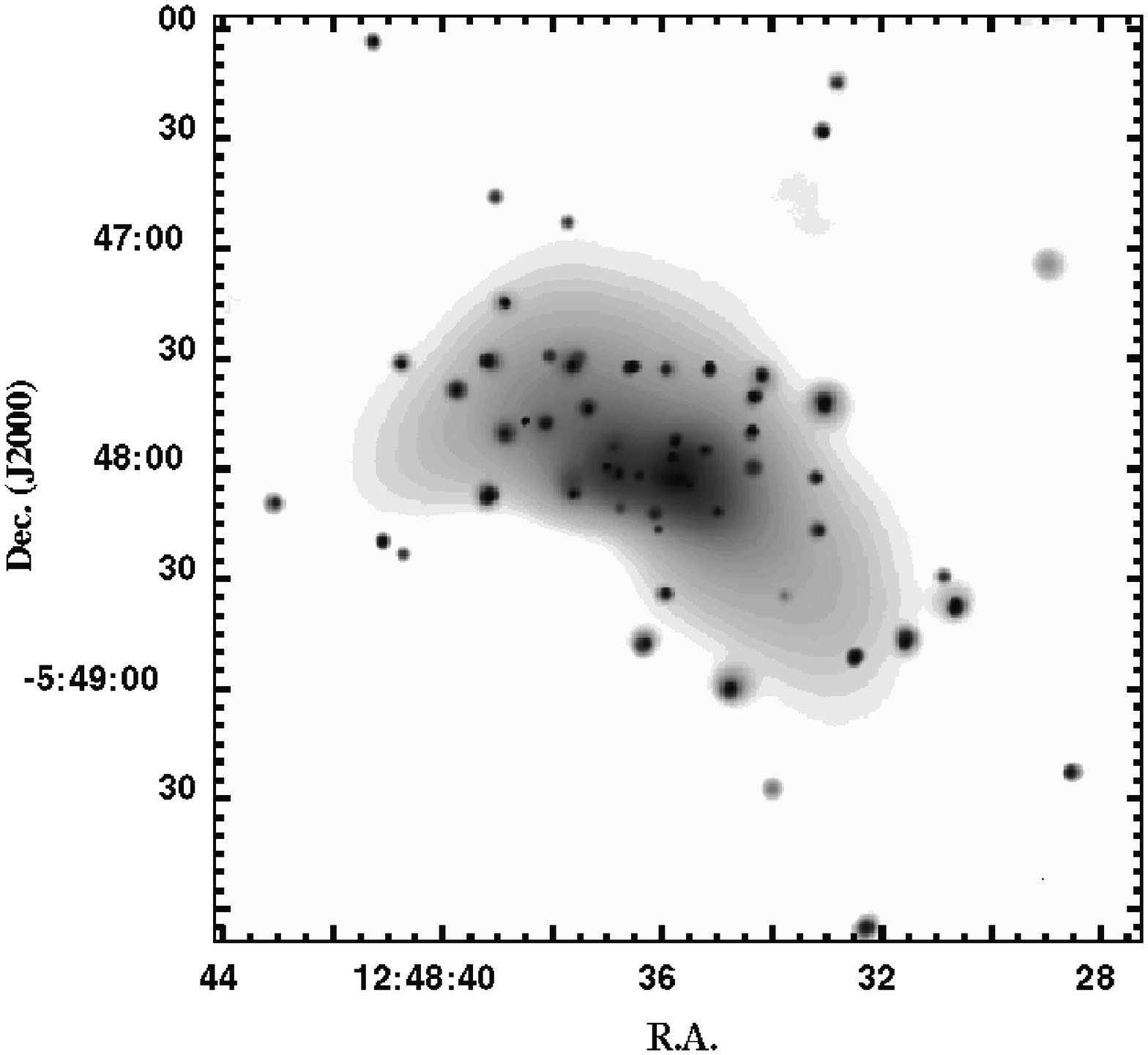}
\figcaption{
The {\it Chandra} X-ray image of a
4\arcmin$\times$4\arcmin\ region around NGC~4697 (0.3--10 keV).
The image was adaptively smoothed to a minimum signal-to-noise
ratio (SNR) of 3 per smoothing beam,
and corrected for background and exposure.
The grey scale is logarithmic, and ranges from $8 \times 10^{-7}$ to
$4 \times 10^{-3}$ cts arcsec$^{-2}$ s$^{-1}$.
\label{fig:xray_mid}}

\vskip0.2truein

A portion of the unresolved emission must also come from LMXBs which
fall below our threshold for reliable detection of resolved sources.
Indeed, if one sets the detection threshold for sources lower, one finds
many more 2-$\sigma$ fluctuations than expected just from Poisson
statistics.
If the observed luminosity function of the resolved sources in
NGC~4697 (Fig.~\ref{fig:xlum} below) is extended down to
$L_X \approx 10^{36}$ ergs s$^{-1}$, the approximate lower limit
for LMXBs in globular clusters (Hertz \& Grindlay 1983),
the contribution of LMXBs to the total band X-ray emission from NGC~4697
increases to 74\% within one effective radius.
The X-ray spectrum of the diffuse emission is somewhat softer than
that of the resolved sources.
If one decomposes the unresolved emission into two components, one with
the same spectrum as that of the sum of the resolved sources plus
a softer component, one finds that the LMXBs would contribute about
77\% of the total band emission.

The remaining $\sim$23\% of the emission would come from a more diffuse
component with a soft ($\sim$0.3 keV) spectrum.
This soft component has a very extended spatial distribution
(Paper II), and is most likely due to interstellar gas.
Additional evidence for ISM emission comes from the adaptively smoothed
X-ray image (Fig.~\ref{fig:xray_mid}).
If the diffuse X-ray emission is from a large number of faint stellar
sources, then it should appear smooth and symmetrical (like the optical 
image).
The diffuse emission in Figure~\ref{fig:xray_mid} is smooth and fairly
symmetric in the inner regions, but there is an extended asymmetrical
feature at the outer regions to the northeast.
Such asymmetries are more easily produced in the distribution of gas
(e.g., by ram pressure) than in the distribution of stellar sources.

\section{Resolved Sources} \label{sec:src}

The discrete X-ray source population on the ACIS S3 image
(Fig.~\ref{fig:xray_whole}) was determined, using a
wavelet detection algorithm with the source detection threshold set at
$10^{-6}$,
which implies that $\la$1 false source
(due to a statistical fluctuation in the background) would be detected
in the entire S3 image.
The minimum detectable flux was about $2.6 \times 10^{-4}$ cts s$^{-1}$
($L_X = 5.0 \times 10^{37}$ ergs s$^{-1}$ at the NGC~4697 distance)
in the 0.3--10 keV band in the part of the chip covered by the galaxy.
Fluxes were corrected for exposure and the instrument PSF.
A total of 90 sources were found;
a complete list with positions and other properties is given
in Paper II.

Some of the detected sources are likely to  be unrelated
foreground or (more likely) background objects.
Based on the source counts in
Brandt et al.\ (2000) and
Mushotzky et al.\ (2000),
we would expect about 10--15 unrelated sources in our observation.
The unrelated sources should mainly be found at larger distances
from the optical center of NGC~4697,
while the sources associated with NGC~4697 should be concentrated to the
center of the galaxy.
Within 2\arcmin\ of the center of NGC 4697 (roughly the region covered
by Fig.~\ref{fig:xray_mid}), $\sim$3 of the $\sim$60 detected
sources would be expected to be unrelated to NGC~4697.

Eight of the X-ray sources coincide (within 1\arcsec) with candidate globular
clusters in NGC~4697.
CXOU J124834.4-055014 (Paper II, Src.~64) is identified with globular
cluster 33 in Hanes (1977),
while CXOU J124846.8-054852 (Paper II, Src.~72, ISB Src.~11) is
associated with Hanes globular cluster 24. 
Six additional X-ray sources
(CXOU J124827.0-054925,
CXOU J124826.1-054729,
CXOU J124830.8-054836,
CXOU J124834.2-054926,
CXOU J124835.9-054551,
CXOU J124841.5-054736)
agree with the positions of candidate globulars 
kindly provided by
Kavelaars (2000).
Given the positional agreement of the X-ray sources with
the globular candidates, it is likely that all 8 of these identifications
are real.
However,
at the distance to NGC~4697, globular clusters are not resolved in
ground-based optical images, and the candidate globulars were identified by
luminosities and (possibly) colors.
As a result, as many as half of them might be unrelated faint
optical objects, rather than globular clusters.

The count rates for the sources were converted into unabsorbed luminosities
(0.3-10 keV) assuming that all of the sources were at the distance of
NGC~4697.
Using the best-fit {\it Chandra} X-ray spectrum  of the resolved sources
(Paper II),
the resulting X-ray luminosities 
range from about $5 \times 10^{37}$ to $2.5 \times 10^{39}$ ergs s$^{-1}$
(calibration uncertainties might affect the luminosities by $\sim$25\%).
The cumulative luminosity function of all of the sources is shown
in Figure~\ref{fig:xlum}.
Note that the luminosity function appears to have a knee
at $L_X \approx 3 \times 10^{38}$ ergs s$^{-1}$.
In fitting the luminosity function, we corrected
for unrelated sources using the deep source counts
(Brandt et al.\ 2000;
Mushotzky et al.\ (2000), assuming that the shape of the
number versus flux relation is that derived from
{\it ROSAT} deep fields
(Hasinger et al.\ 1998).
A single power-law (plus the background contribution) fit
to the luminosity function could be rejected
at the $>$95\% confidence level.
However, a broken power-law fit was successful:
\begin{equation} \label{eq:xlum}
\frac{ d N }{ d L_{38} } = N_o \,
\left( \frac{ L_{38} }{ L_b } \right)^{-\alpha} \,  ,
\end{equation}
with $\alpha = \alpha_l$ for $L_{38} \le L_b$ and 
$\alpha = \alpha_h$ for $L_{38} > L_b$.
Here, $L_{38}$ is the X-ray luminosity (0.3--10 keV) in units
of $10^{38}$ ergs s$^{-1}$.
The best fit, determined by the maximum-likelihood method, gave
$N_o = 8.0^{+7.1}_{-5.2}$,
$\alpha_l = 1.29^{+0.36}_{-0.49}$, 
$\alpha_h = 2.76^{+1.81}_{-0.39}$,
and a break luminosity of
$L_b = 3.2^{+2.0}_{-0.8} \times 10^{38}$ ergs s$^{-1}$
(Figure~\ref{fig:xlum}).
The errors (at the 90\% confidence level) were determined by
Monte Carlo simulations.

%
%
\centerline{\null}
\vskip2.00truein
\includegraphics{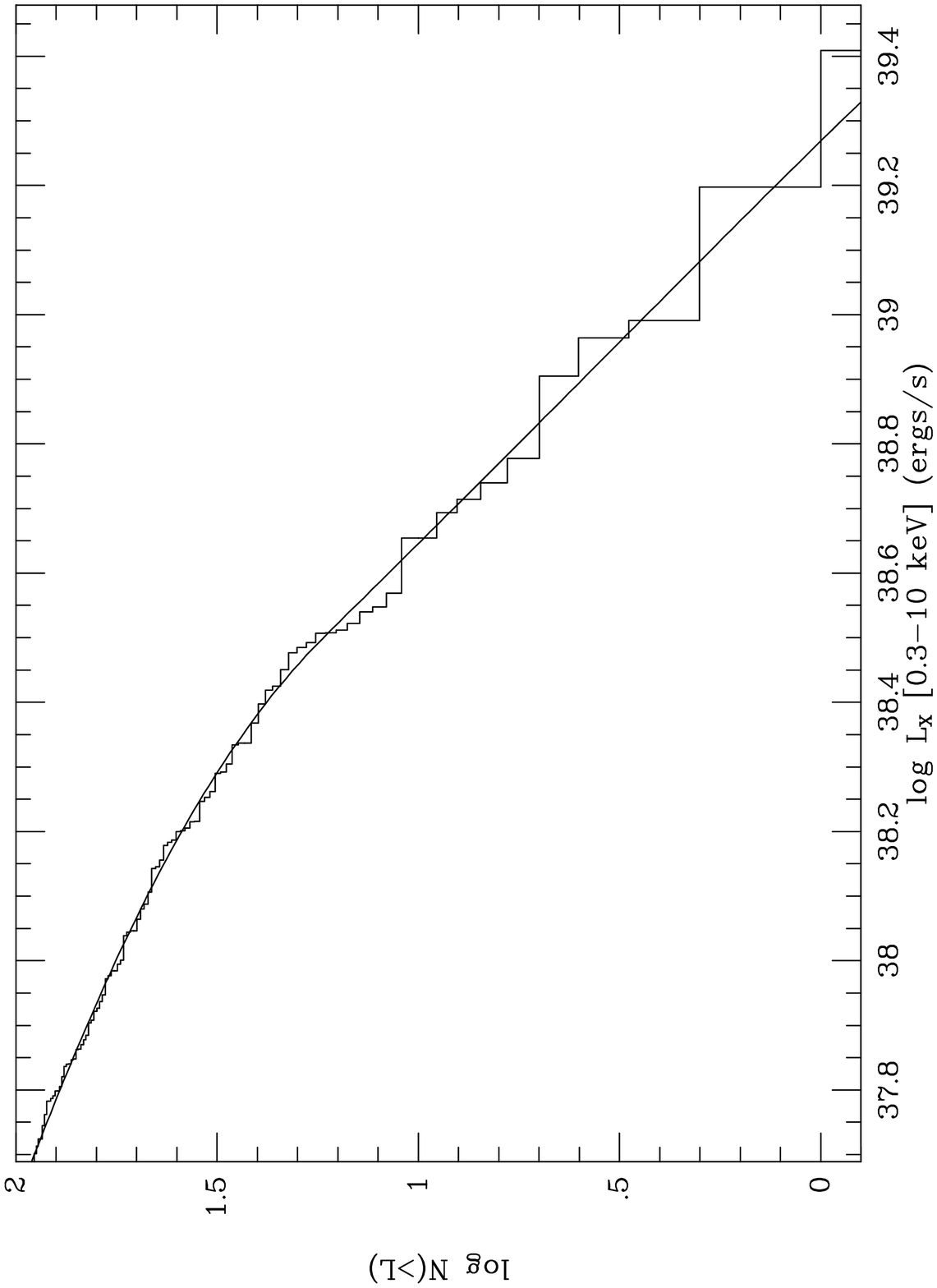}
\figcaption{
The histogram is the observed cumulative X-ray luminosity function of all
resolved sources within the {\it Chandra} S3 field.
The continuous curve is the sum of the best-fit LMXB luminosity function
for NGC~4697 (eq.~\protect\ref{eq:xlum}), plus the expected background
sources counts.
\label{fig:xlum}}

\vskip0.2truein

The break luminosity is similar to the Eddington luminosity
for spherical accretion onto a 1.4 $M_\odot$ neutron star.
This suggests that the sources with luminosities above this
break are accreting black holes, while those below the break
are predominantly neutrons stars.
This would imply that NGC~4697 contains $\ga$15 luminous
X-ray binary systems containing black holes.
If the more luminous of these systems ($L_X \sim 10^{39}$ ergs s$^{-1}$)
are limited by the Eddington luminosity, they must contain fairly massive
($M \ga 8 \, M_\odot$) black holes.

We studied the crude spectral properties of the resolved sources by
using hardness ratios
(full results and spectral analysis in Paper II).
Hardness ratios or X-ray colors have the advantage that they can
be applied to weaker sources.
Figure~\ref{fig:colors_noerr} plots H31 vs.\ H21.
For comparison, the hardness ratios (H21,H31) are
$(-0.38,-0.57)$ for all of the emission, 
$(-0.69,-0.82)$ for the unresolved emission,  and
$(-0.14,-0.37)$ for the sum of the sources, all within one effective
radius.

There are three moderate luminosity sources with hardness ratios of
$(-1,-1)$, 
which means that they have no detectable emission beyond 1 keV.
These three sources are almost certainly supersoft sources
(e.g., Kahabka \& van den Heuvel 1997).
There are three sources, one of which is the brightest source in the field,
with values of $\sim$$(1,1)$, all of which are located $>$3\farcm5 
from the center of NGC~4697.
These are probably unrelated, strongly absorbed AGNs.
There are eight sources near $(0,-1)$ which have essentially no hard
emission.
Six of these sources are at large radii ($>$2\farcm2),
which suggests that this population is also unrelated to NGC~4697.
Most of the sources lie in a diagonal swath centered at about
$(-0.15,-0.40)$.
These values are similar to but slightly harder than the integrated colors
for the entire galaxy, but considerably harder than the values for the
unresolved emission.
The X-ray sources identified with globular cluster candidates
have hardness ratios which are similar to other LMXBs, with the
exception of CXOU J124834.4-055014, whose hardness ratios suggest
that it is actually a background AGN.

%
%
\centerline{\null}
\vskip2.80truein
\includegraphics{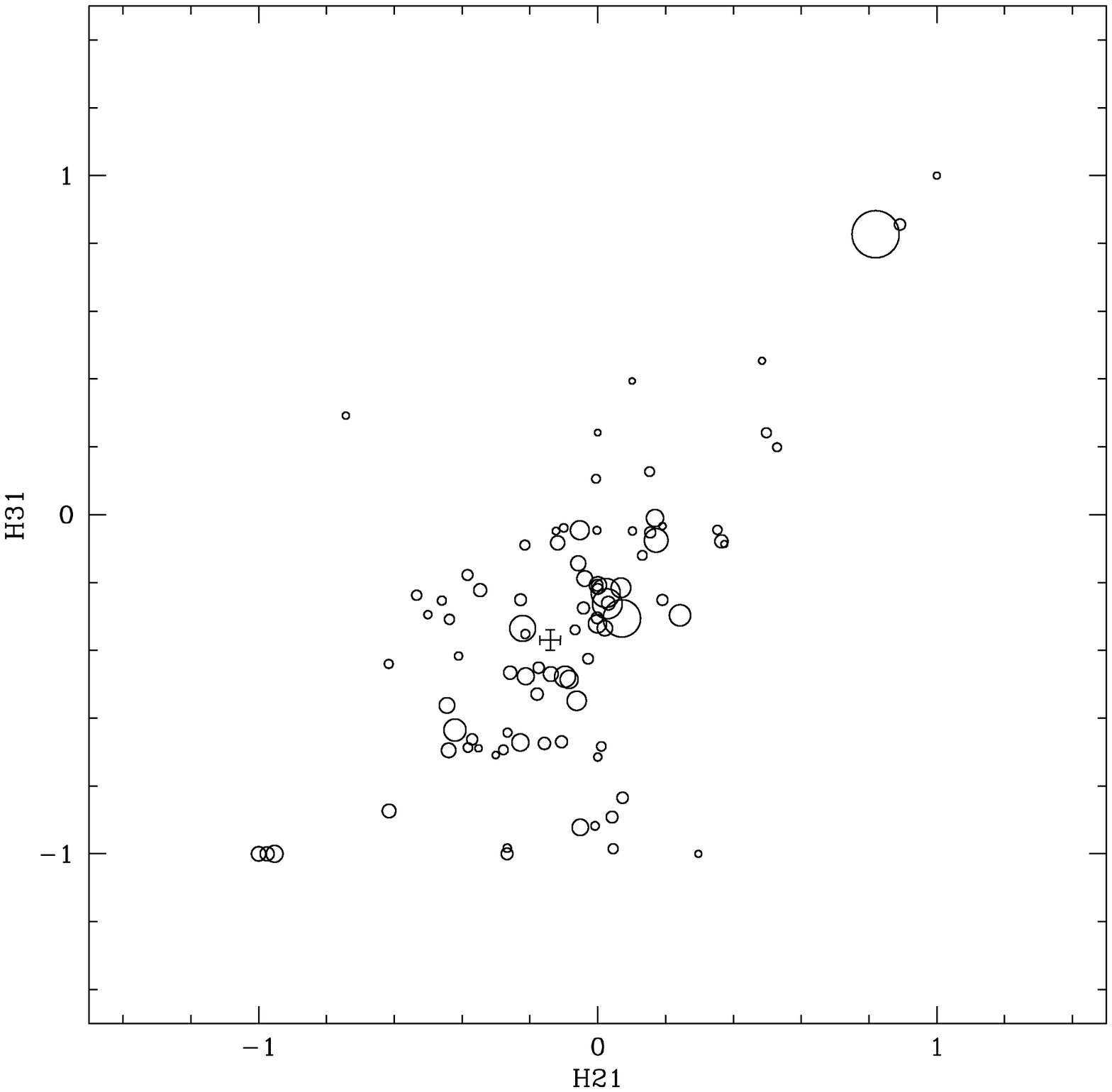}
\figcaption{
The hardness ratios for the NGC~4697 sources.
Here, $H21 \equiv ( M - S ) / ( M + S )$
and $H31 \equiv ( H - S ) / ( H + S ) $, where $S$, $M$, and $H$ are
the counts in our soft (0.3--1 keV), medium (1--2 keV), and hard
(2--10 keV) bands, respectively.
The area of the circle for each source is proportional to its flux.
The three sources near $(-1,-1)$ are supersoft sources.
The 11 sources near $(1,1)$ and $(0,-1)$ may be unrelated background
objects.
The cross indicates the mean hardness for all of the sources.
\label{fig:colors_noerr}}

\vskip0.2truein

\section{Conclusions} \label{sec:conclude}

Our high spatial resolution {\it Chandra} observation of the X-ray faint
elliptical galaxy NGC~4697 resolves most of the X-ray emission
(61\% within one effective radius) into point sources.
A total of 90 individual sources are detected, of which $\sim$80
are low mass X-ray binaries (LMXBs) associated with NGC~4697.
Much of the emission is resolved even in the softest band.
The dominance of LMXBs implies that this and other X-ray faint ellipticals
have lost most of their interstellar gas.

Of the remaining unresolved emission, it is likely that about half is
from fainter LMXBs, while about 23\% of the total emission is probably
from interstellar gas with
$kT \sim 0.3$ keV.
Faint, diffuse X-ray emission in the {\it Chandra} image is asymmetrical
and very extended,
which also suggests that a portion of the emission is due to hot gas.
Ram pressure may have produced the asymmetric feature.

Three of the resolved sources appear to be supersoft sources.
Eight of the resolved sources in the outer parts of NGC~4697 (about 25\%)
are coincident with candidate globular clusters.
On the other hand, the candidate globular clusters contain about
0.1\% of the optical light of the galaxy in this region.
This indicates that globulars have an unusually large fraction
of X-ray binaries (by a factor of $\sim$250), as is seen in our Galaxy
(e.g., Hertz \& Grindlay 1983).

The X-ray luminosities (0.3--10 keV) of the resolved LMXBs range from
$\sim$$5 \times 10^{37}$ to $\sim$$2.5 \times 10^{39}$ ergs s$^{-1}$.
The luminosity function has a ``knee'' at
$3.2 \times 10^{38}$ ergs s$^{-1}$, which is roughly the Eddington
luminosity of a 1.4 $M_\odot$ neutron star (NS).
This knee might provide a standard candle which could be used to
determine distances to galaxies.
This knee may separate accreting NS and
black hole (BH) binaries.
If the brightest sources in NGC~4697 are Eddington limited, they must
contain fairly massive BHs.

Our detection of a large population of binaries with NSs and massive BHs
provides perhaps the most direct evidence that this elliptical
galaxy (or its progenitors) once contained a large number of massive
main sequence stars.
The population of LMXBs provides a tool to study the high mass end
of the initial mass function of early-type galaxies, long after the
massive main sequence stars have died.
The detection of NS and massive black holes in NGC~4697 also provides
the first evidence for the current existence of massive stars
(albeit degenerate) in a normal, optically luminous elliptical galaxy.

\acknowledgements
We are extremely grateful to JJ Kavelaars for providing his unpublished list
of globular clusters in NGC~4697.
Bill Harris, JJ Kavelaars, and  Arunav Kundu also gave helpful comments
on the globular cluster population of NGC~4697.
Support for this work was provided by the National Aeronautics and Space
Administration through Chandra Award Number GO0-1019X issued by
the Chandra X-ray Observatory Center, which is operated by the Smithsonian
Astrophysical Observatory for and on behalf of NASA under contract
NAS8-39073.
J. A. I. was supported by {\it Chandra} Fellowship grant PF9-10009, awarded
through the {\it Chandra} Science Center.


\begin{references}

\reference{}
Brandt, W. N., et al., 2000, ApJ, 199, 2349

\reference{}
Fabbiano, G., Kim, D.-W., \& Trinchieri, G. 1994, ApJ, 429, 94

\reference{}
Faber, S. M., Wegner, G., Burstein, D., Davies, R. L., Dressler, A.,
Lynden-Bell, D., \& Terlevich, R. J. 1989, ApJS, 69, 763

\reference{}
Forman, W., Jones, C., \& Tucker W. C. 1985, ApJ, 293, 102

\reference{}
Hanes, D. A. 1977, MmRAS, 84, 45

\reference{}
Hasinger, G., Burg, R., Giacconi, R., Schmidt, M., Tr\"umper, J., \&
Zamorani, G. 1998, A\&A, 329, 495

\reference{}
Hertz, P., \& Grindlay, J. E. 1983, ApJ, 275, 105

\reference{}
Irwin, J. A., \& Sarazin, C. L. 1998a, ApJ, 494, L33

\reference{}
Irwin, J. A., \& Sarazin, C. L. 1998b, ApJ, 499, 650

\reference{}
Irwin, J. A., Sarazin, C. L., \& Bregman, J. N. 2000, ApJ, 544, in press (ISB)

\reference{}
Kahabka, P., \& van den Heuvel E. P. J. 1997, ARAA 35, 69 

\reference{}
Kavelaars. J. J. 2000, private communication

\reference{}
Kim, D. -W., Fabbiano, G., Matsumoto, H., Koyama, K., \& Trinchieri, G.
1996, ApJ, 468, 175

\reference{}
Matsumoto, H., Koyama, K., Awaki, H., \& Tsuru, T., Loewenstein, M.,
\& Matsushita, K. 1997, ApJ, 482, 133

\reference{}
Monet D., et al.,
1998, USNO-A V2.0, A Catalog of Astrometric Standards
(Flagstaff: U.S.\ Naval Observatory)

\reference{}
Mushotzky, R. F., Cowie, L. L., Barger, A. J., \& Arnaud, K. A. 2000,
Nature, 404, 459

\reference{}
Pellegrini, S. 1994, A\&A, 292, 395

\reference{}
Pellegrini, S., \& Fabbiano, G. 1994, ApJ, 429, 105

\reference{}
Sarazin, C. L., Irwin, J. A., \& Bregman, J. N. 2000, in preparation
(Paper II)

\end{references}
\end{document}